\documentclass[12pt,dvips]{article}
\usepackage{epsfig}

\newcommand{\be}{\begin{equation}}
\newcommand{\ee}{\end{equation}}

\begin{document}

\thispagestyle{empty}
\date{\today}
\title{
\vspace{-5.0cm}
\begin{flushright}
{\normalsize UNIGRAZ-UTP-17-09-99}\\
{\normalsize BUTP-99/18}\\
\end{flushright}
\vspace*{2cm}
Effects of Topology in the Dirac Spectrum of Staggered Fermions
\thanks{Supported by Fonds 
zur F\"orderung der Wissenschaftlichen Forschung 
in \"Osterreich, Project P11502-PHY.} }
\author
{\bf F. Farchioni \\  \\ 
Institut f\"ur Theoretische Physik,\\
Universit\"at Bern, CH-3012 Bern, Switzerland \\  \\
\bf I. Hip and C. B. Lang  \\  \\
Institut f\"ur Theoretische Physik,\\
Universit\"at Graz, A-8010 Graz, Austria}
\maketitle
\begin{abstract}
We compare the lower edge spectral fluctuations  of the staggered lattice Dirac
operator for the Schwinger model  with the predictions of chiral Random Matrix
Theory (chRMT). We verify their range of applicability,  checking in particular
the r\^ole of non-trivial topological sectors and the flavor symmetry of the
staggered fermions for finite lattice spacing. Approaching the continuum limit
we indeed find clear signals for topological  modes in the eigenvalue spectrum.
These findings indicate problems in the  verification of the chRMT predictions.
\end{abstract}


\newpage

\section{Introduction}
\label{sec:intro}

Leutwyler and  Smilga's sum rules \cite{LeSm92} for the eigenvalues  of the
chiral Dirac operator for finite volume QCD suggest that the
statistical fluctuations of the spectrum of this operator are universal in
the infrared region;  these should be described  by a chiral Random Matrix
Theory (chRMT) \cite{ShVe93}. (For a recent review  on Random Matrix Theory
in general cf. \cite{GuMuWe98}.)  Universality has been proved extensively
in chRMT; for the case of QCD, see \cite{DaAkOsDoVe}. Explicit lattice
calculations may provide an answer to the question, whether QCD lies really
inside one of the  universality classes of chRMT.

Chiral symmetry, which is the main ingredient here, may be effectively
realized on the lattice by the so-called  ``overlap fermions'' \cite{NaNe93}
and by the Fixed-Point (FP) fermion action \cite{Ha98a}, both satisfying the
Ginsparg-Wilson \cite{GiWi82} condition.  Ginsparg-Wilson fermions -- in
spite of their theoretical appeal --  present however non-trivial technical
difficulties in a dynamical set-up (for first experiences in this context,
see \cite{FaHiLaWo99, EdHeNa99}).

Kogut-Susskind ``staggered'' fermions, which are the object of this letter,
realize chirality only partially, but are also much simpler, and still
represent the typical framework for lattice QCD calculations  when chirality
is relevant. However, this is  the main conclusion of this letter,
comparison with chRMT is in this case  not straightforward and subject to
additional restrictions (besides the usual ones well-known in the
literature) applying as long as the lattice cut-off is finite.

A class of problems comes from topology:   when comparing lattice eigenvalue
spectra with chRMT predictions, the validity of the index theorem
\cite{AtSi71} for the Dirac operator is assumed, while the staggered Dirac
operator has no exact zero modes and a sound fermionic definition of  the
topological charge is not possible. Another class of problems is due to the
flavor structure of staggered fermions, which is different at finite cut-off
and in the continuum limit. Each copy of staggered fermions describes in the
continuum $N_f=2^{d/2}$ degenerate massless physical fermions;  at finite 
cut-off the full chiral symmetry is broken into an abelian subgroup
resembling the non-degenerate ($N_f=1$) situation, with the relevant
difference that the symmetry is here {\em anomaly-free}.

We study these issues in the simplified framework of the Schwinger model in
the quenched and the dynamical situation. A particularity of two dimensions is
that a continuous  symmetry cannot be broken by the vacuum \cite{MeWaCo66}.
Since the chiral symmetry at finite cut-off is non-anomalous though abelian,
the usual fermion condensate vanishes  even if staggered fermions act
as just one flavor when comparing Dirac eigenvalue spectra with chRMT. With
vanishing condensate, a different kind of universality  should apply in the
infrared part of the spectrum \cite{JaNoPaZaBrHi98}.

In a Monte Carlo simulation, all this applies of course only for the
unquenched  (dynamical) data. For the ``quenched'' model (corresponding to
$N_f=0$)  the usual chRMT universality is expected to hold
\cite{FaHiLaWo99}. In this case, however, different complications come into
play when the thermodynamic limit of the theory is taken \cite{Sm92}. Some
results concerning the quenched data  have been discussed in
\cite{FaHiLa99c}.

\section{Predictions from Random Matrix Theory}
\label{sec:rmt}

The lower edge of the spectrum of the Dirac operator   (i.e. its infra-red
region) is expected to be universal at the scale of the level spacing
$\Delta\lambda$ (microscopic scale) $\sim\,1/V$.  The `microscopic' scaling
variable is $z=V\,\Sigma\,\lambda$, where 
\be
\label{eq:rho0}
\Sigma\equiv\pi\lim_{\lambda\to 0}\lim_{V\to\infty}\,\rho(\lambda)\;,
\ee
and $\rho(\lambda)$ denotes the associated  spectral density per unit volume. 
When  comparing the statistical properties of the Dirac operator with chRMT,
$\Sigma$  is an external scale parameter related to the dynamics of the
physical system  under study. The Banks-Casher formula \cite{BaCa80} implies 
$\Sigma=-\langle\bar{\psi}\psi\rangle$.

Three classes of universality are predicted by chRMT \cite{Ve94}, corresponding
to (chiral) orthogonal,  unitary and symplectic ensembles (chOE, chUE and
chSE respectively). The universal properties considered in this letter are the
microscopic spectral density
\be\label{eq:sp_dens}
\rho_s(z) = \lim_{V\to\infty}\frac{1}{\Sigma}\,
\rho\left (\frac{z}{V\Sigma}\right)
\ee
and the probability distribution of the  smallest eigenvalue $P(\lambda_{\rm
min})$. These distributions depend on the topological charge $\nu$ of the
background gauge configuration and on the number of flavors $N_f$ of quarks
included in the dynamics. We also checked the level spacing statistics
\cite{FaHiLa99c}.

\section{Statistical properties of the spectrum}
\label{sec:res_edge}

Staggered fermions partially realize chiral symmetry on the lattice. One
species  of staggered fermion describes $N_f=2^{d/2}$ Dirac fermions in the
continuum limit. For finite lattice spacing the chiral symmetry
$U_V(2^{d/2})\times U_A(2^{d/2})$  is however broken down to a $U_V(1)\times
U_A(1)$ {\em non-anomalous} sub-symmetry, which accounts for the 
reflection-invariance  for the purely imaginary spectrum.

In chRMT, the number of dynamical fermions $N_f$  and the topological charge
$\nu$ of the background configuration enter as external parameters. Our
(compact) gauge configurations were sampled in the quenched set-up, i.e.
according to the pure gauge measure (given by the standard  plaquette action).
This corresponds in chRMT to the case $N_f=0$. Assuming no extra symmetry for
the Dirac operator (besides the one mentioned above)  in our two-dimensional
set-up, the statistical fluctuations of the spectrum    in the infrared region
are expected \cite{Ve94} to be described by chUE; this  was confirmed by
previous results \cite{FaHiLaWo99} with the Fixed Point  and Neuberger's
overlap action, where however, deviations (i.e.  a distribution resembling
chSE) were observed whenever the physical volume was too small.

In order to explore the dynamics of fermions, we then re-weighted the quenched
configurations with the fermion determinant. For $d=4$ this procedure usually
introduces  large statistical fluctuations; in our two-dimensional model these
fluctuations did not appear to be too harmful \cite{FaHiLaWo99}. Since the
chiral symmetry of lattice staggered fermions  is abelian, one expects
\cite{BeGoMe99} that  the dynamical situation is described by a chRMT with
$N_f=1$. The two-dimensional situation is however peculiar due to the
Mermin-Wagner-Coleman theorem stating that a continuous symmetry cannot be
broken by the vacuum; only an anomaly can do that.  Since the operator
$\bar{\psi}(x)\psi(x)$ is not invariant under the residual (non-anomalous)
abelian chiral symmetry of staggered fermions,  its expectation value vanishes
\cite{BoKo86}. (One can argue \cite{GaSe94} that the simplest observable
signaling the  breaking of the {\em anomalous} axial symmetry    is in this
context $\langle\bar u u \bar d d\rangle$.) With vanishing fermion condensate,
the universality class of chUE does not apply anymore,  and a new kind of
universality holds \cite{JaNoPaZaBrHi98}.

The validity of the $N_f=1$ chRMT predictions is restricted. In the continuum
limit the irrelevant terms of the action  breaking the full symmetry vanish
and a transition to $N_f=2$  is due to occur with doubling of the spectrum. In
this limit, two degenerate fermions described by two-component  spinors are
the effective degrees of freedom,  the two-flavor (massless) Schwinger model
being the underlying  continuum theory \cite{BoKo86}.

RMT predictions depend also on the topological charge of the background gauge
configuration. In the continuum the topological charge is related  to the
index of the Dirac operator \cite{AtSi71} (the number of zero modes counted
according to their chirality). On the lattice, a consistent definition of the
topological charge in terms of the index of the Dirac operator is possible
only for Ginsparg-Wilson (i.e. Overlap and FP) fermions \cite{HaLaNi98}.  For
staggered fermions, away from the continuum limit, the Dirac operator has no
exact zero modes, and the naive expectation is that all configurations show
spectral distributions described by chRMT predictions for the trivial sector 
\cite{BeGoMe99,DaHeNi99}.  

However, for lattices that are fine enough one should attain the continuum
situation. We will explore this hypothetical scenario and check possible
systematic effects when comparing the spectrum of the Dirac operator with
chRMT predictions. Lacking (in this case) a consistent fermionic definition 
of the topological charge $\nu$, we rely on the geometric definition of the
gauge configurations. In the analysis we distinguish between the sectors with
trivial ($\nu=0$) and non-trivial topological charge.

\subsection{The simulation}

We considered statistically independent gauge configurations for lattices of
size $16^2$ ($\beta$ = 2, 4, 6; 10000 configurations); $24^2$ ($\beta$ = 2, 4,
6, 8 ; 5000 configurations) and $32^2$ ( $\beta$ = 2, 4, 6; 5000 configurations).
Even-odd preconditioning and standard LAPACK routines were used to diagonalize 
the staggered Dirac operator.

\begin{figure}[t]
\begin{center}
\epsfig{file=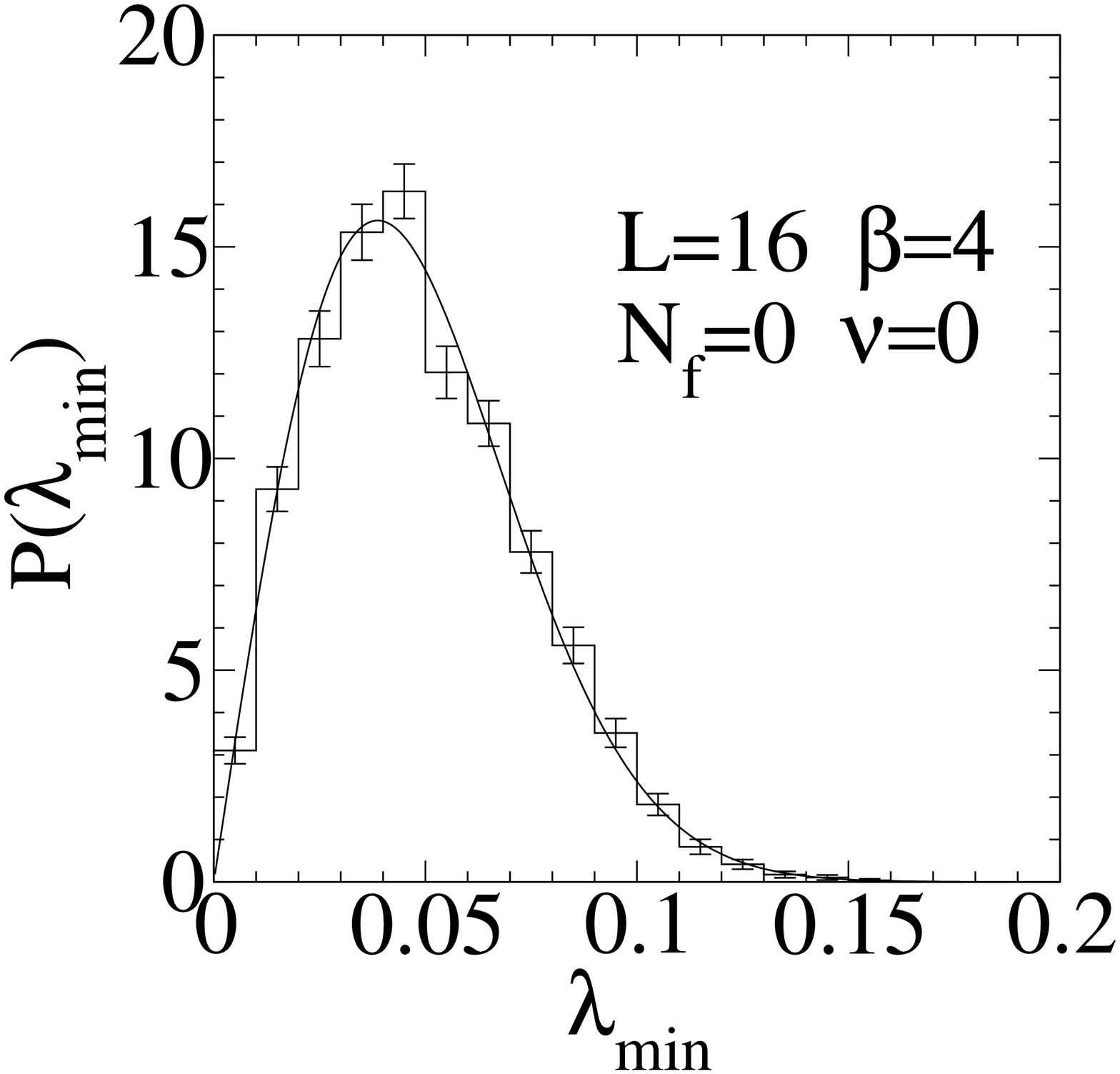,width=4 truecm}
\epsfig{file=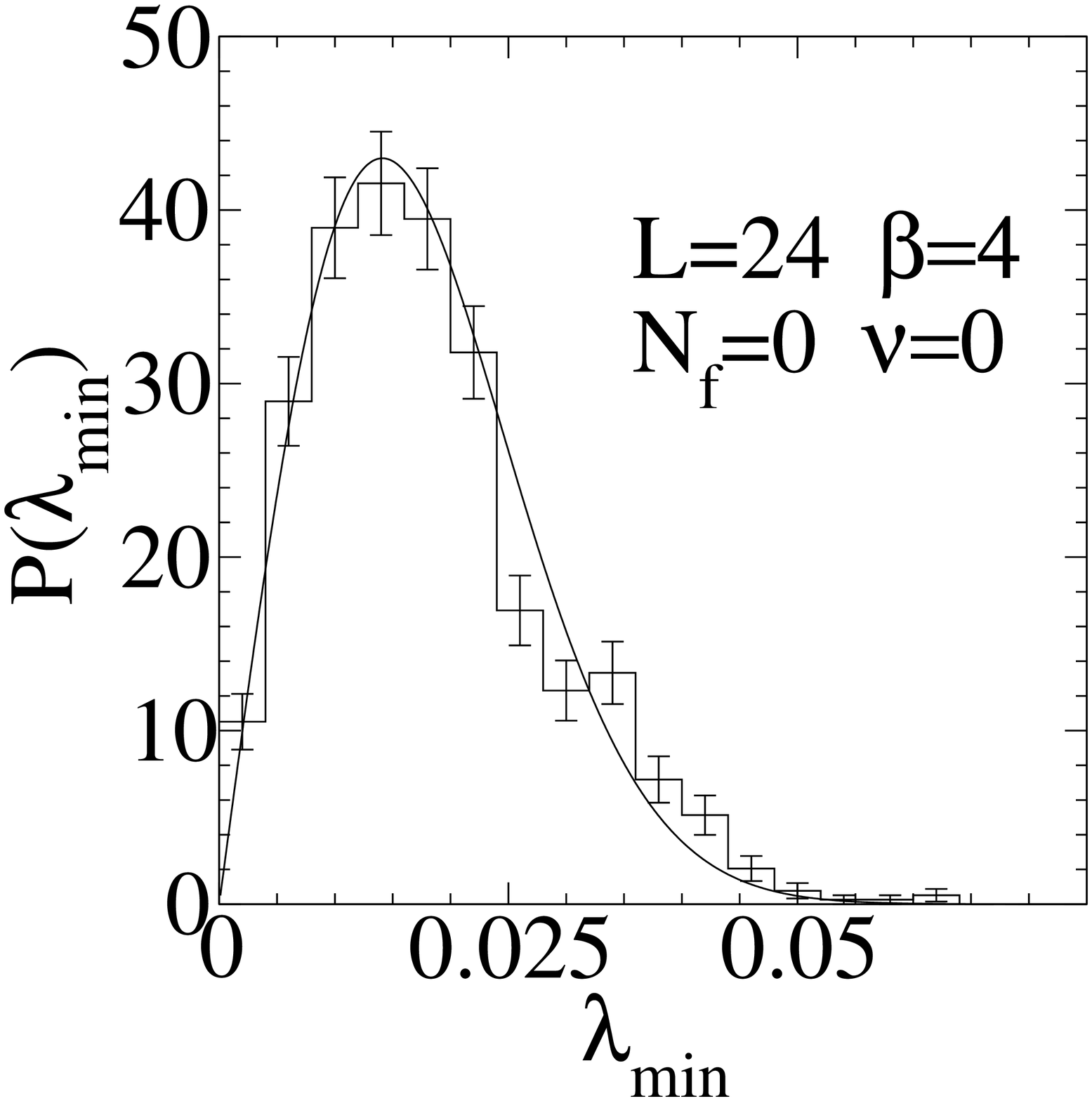,width=4 truecm}
\epsfig{file=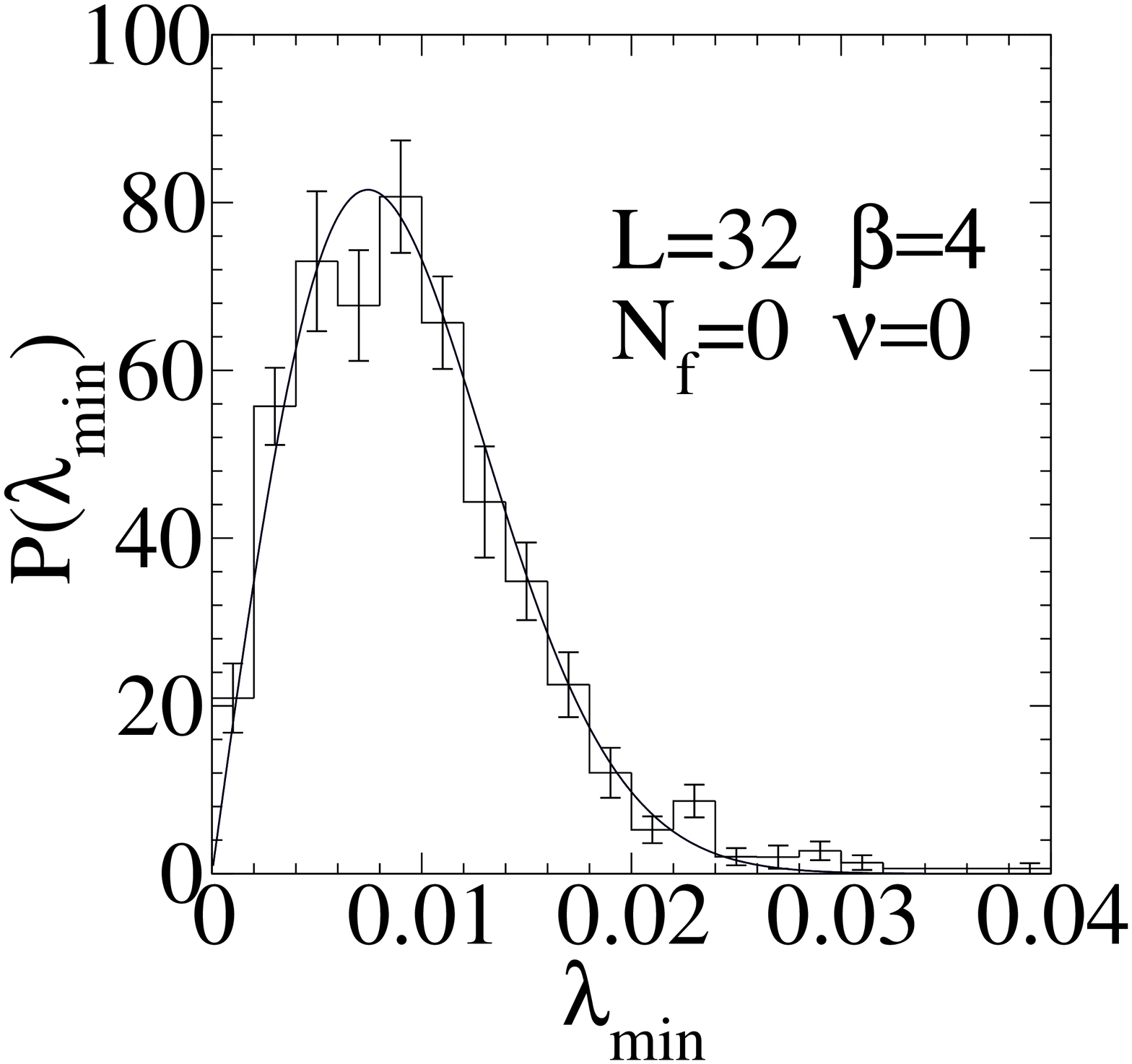,width=4 truecm}
\end{center}
\caption{\label{fig:min}
Distribution of the smallest eigenvalue for  different lattice sizes and $\beta=4$
(quenched case); the continuous curve represents the theoretical  prediction of
chUE.}
\end{figure}

\begin{figure}[tp]
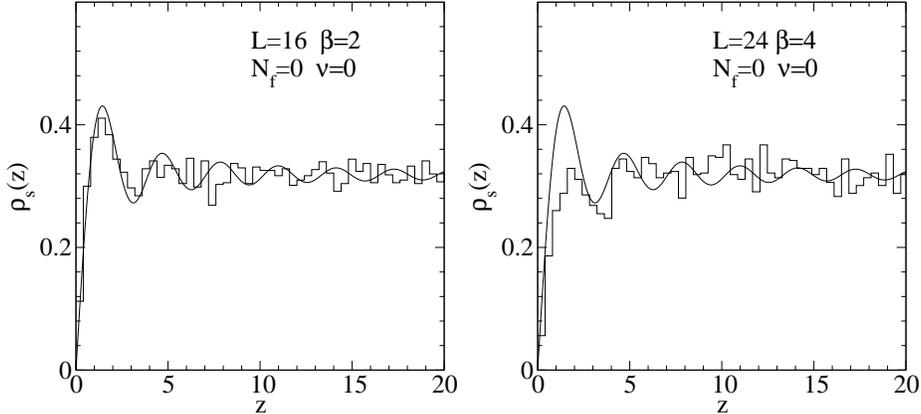

\begin{center}
\epsfig{file=figs/histo_l16_b2.eps,width=6 truecm}
\epsfig{file=figs/histo_l24_b4.eps,width=6 truecm}
\end{center}
\caption{\label{fig:dens}
Microscopic spectral density $\rho_s(z)$ for 
different lattices and values of $\beta$ (quenched case);
the continuous curves represent the theoretical 
prediction of chUE.}
\end{figure}

\paragraph{The trivial sector.}
Fig. \ref{fig:min}  shows the  distribution of the smallest eigenvalue
$P(\lambda_{\rm min})$ for some lattice volumes.  The histograms are obtained
by choosing configurations with geometrical topological charge $\nu=0$; this
reduces the statistics to $30\% - 10\%$ of the original one depending on $L$
and $\beta$.

We observe consistency with chUE except for $L=16$, $\beta\ge 6$, where the
physical volume is likely to be too small for chRMT to apply.  For the fixed
point and Neuberger's action an apparent  transition to chSE,  not observed
here, took place for small lattices and/or high values of  $\beta$
\cite{FaHiLaWo99}.

\begin{figure}[t]
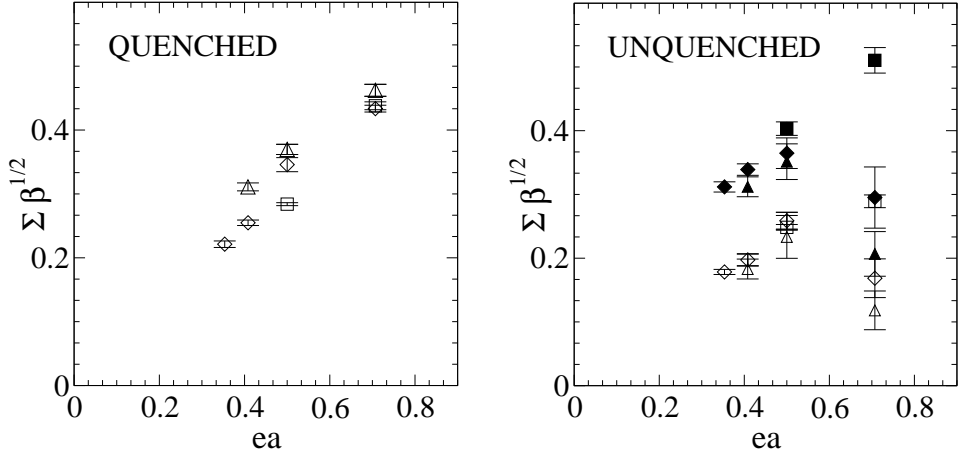

\begin{center}
\epsfig{file=figs/cond.eps,width=6 truecm}\hspace{5mm}
\epsfig{file=figs/cond_unq.eps,width=6 truecm}
\end{center}
\caption{\label{fig:sigma}
The condensate $\Sigma$ as obtained from the  fits on the smallest eigenvalue
distribution using chRMT predictions, for the quenched (left) and re-weighted
(right) gauge configurations:  $L=16$ (boxes), $L$=24 (diamonds), $L$=32
(triangles);  in the dynamical case we plot the fit values obtained from
comparison with chUE  for $N_f=1$ (open symbols) and $N_f=2$ (full symbols).
}
\end{figure}

The fit of the chUE distribution provides an estimate for $\Sigma$ which is
given in Table \ref{tab:sigma} and in Fig. \ref{fig:sigma}, left (we are using
$e\,a=1/\sqrt{\beta}$ as abscissa,  in the ordinate we report 
$\Sigma\sqrt{\beta}=-\langle\bar{\psi}\psi\rangle_{\rm cont}/e$; the equations
hold in the weak coupling limit of the Schwinger model with one flavor).

Following RMT the spectrum is to be separated in a fluctuation part
(conjectured to follow chRMT predictions) and a smooth background. The
information on the chiral condensate is contained in both. Removing the smooth
background is done by ``unfolding'': the eigenvalues are mapped to  a variable
$z$, such that the {\em average} level spacing is constant. In Fig.
\ref{fig:dens} we plot the microscopic spectral density  $\rho_s(z)$ for the
unfolded eigenvalues for two situations. In the unfolding  process the average
value of the spectral density near $\lambda=0$ provides a direct estimate for
$\Sigma$ (see (\ref{eq:rho0})).  Neglecting finite-size effects, this value
should  agree with the number obtained from the smallest eigenvalue
distribution in the scaling region of chRMT. We find discrepancies  between
the two determinations for $\beta\ge 4$,  the ``background'' value of $\Sigma$
(i.e. $\pi\rho(0)$) coming out larger; as a consequence of this discrepancy, 
the spectral density does not follow the predicted shape anymore (Fig.
\ref{fig:dens}, right). We interpret this as due to the  transition to the
full flavor-symmetric situation of the continuum limit, which in the quenched
case implies trivial doubling of the spectrum.

\begin{figure}[tp]
\begin{center}
\epsfig{file=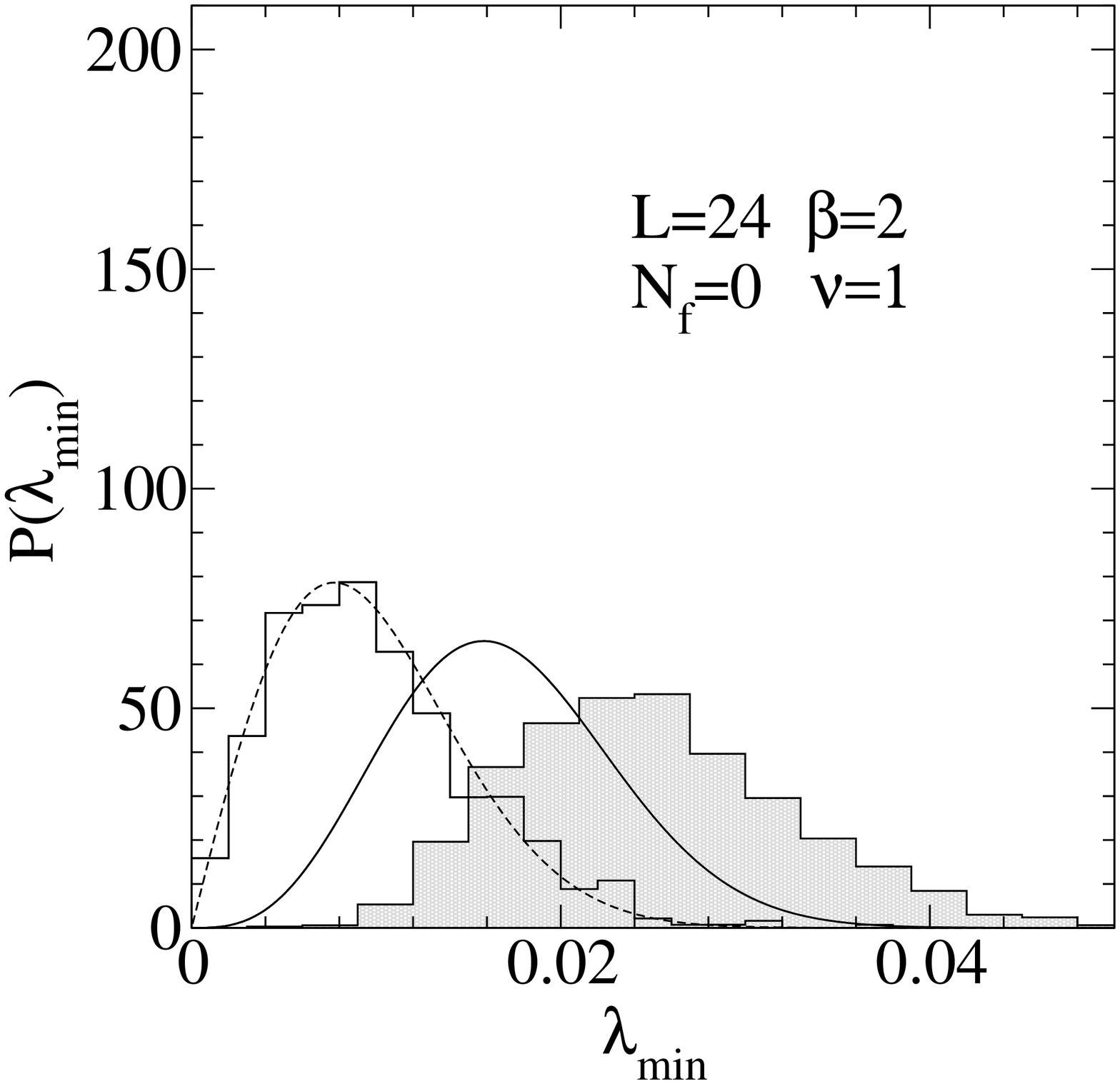,width=6.5 truecm}
\epsfig{file=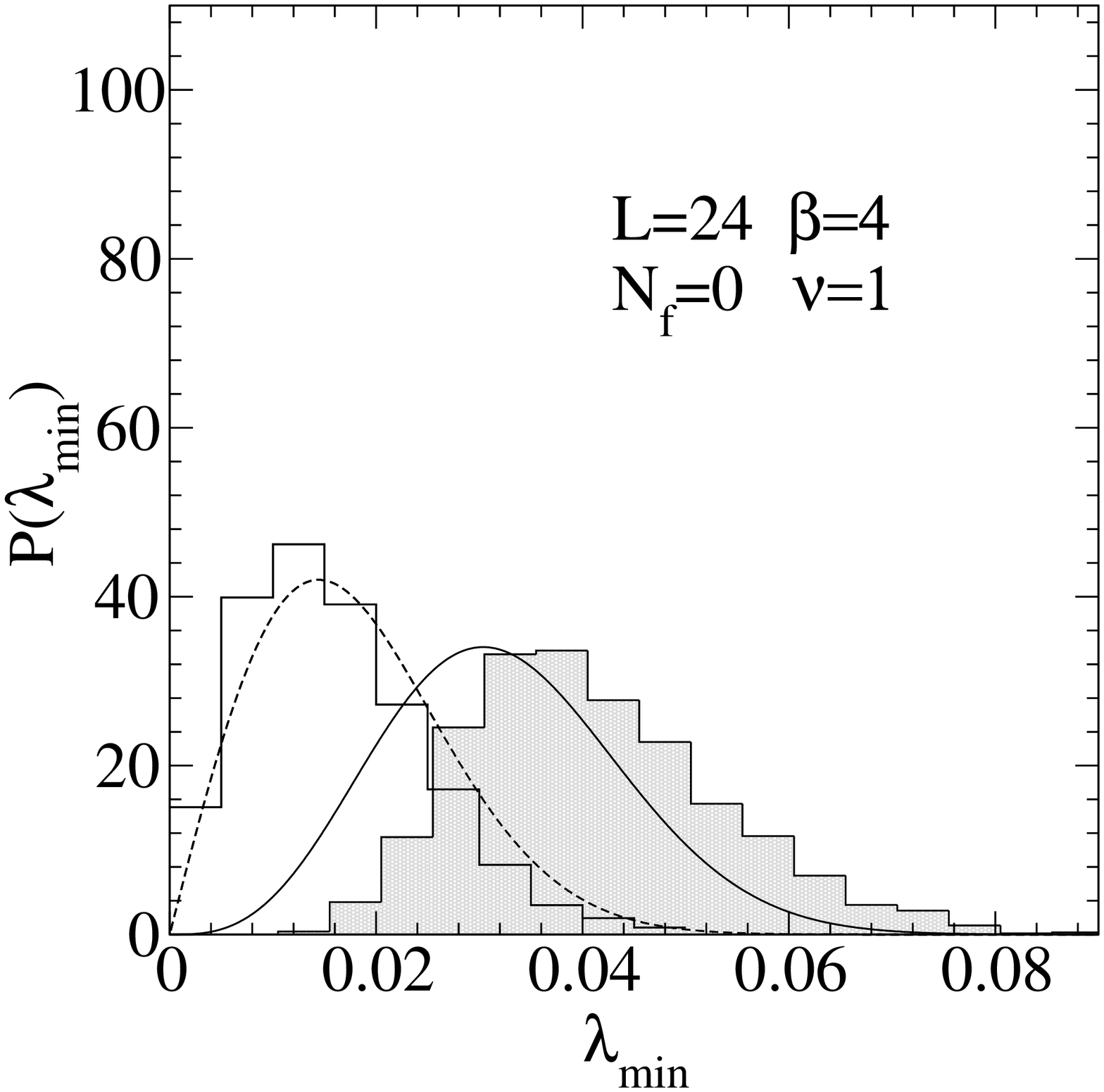,width=6.5 truecm}
\epsfig{file=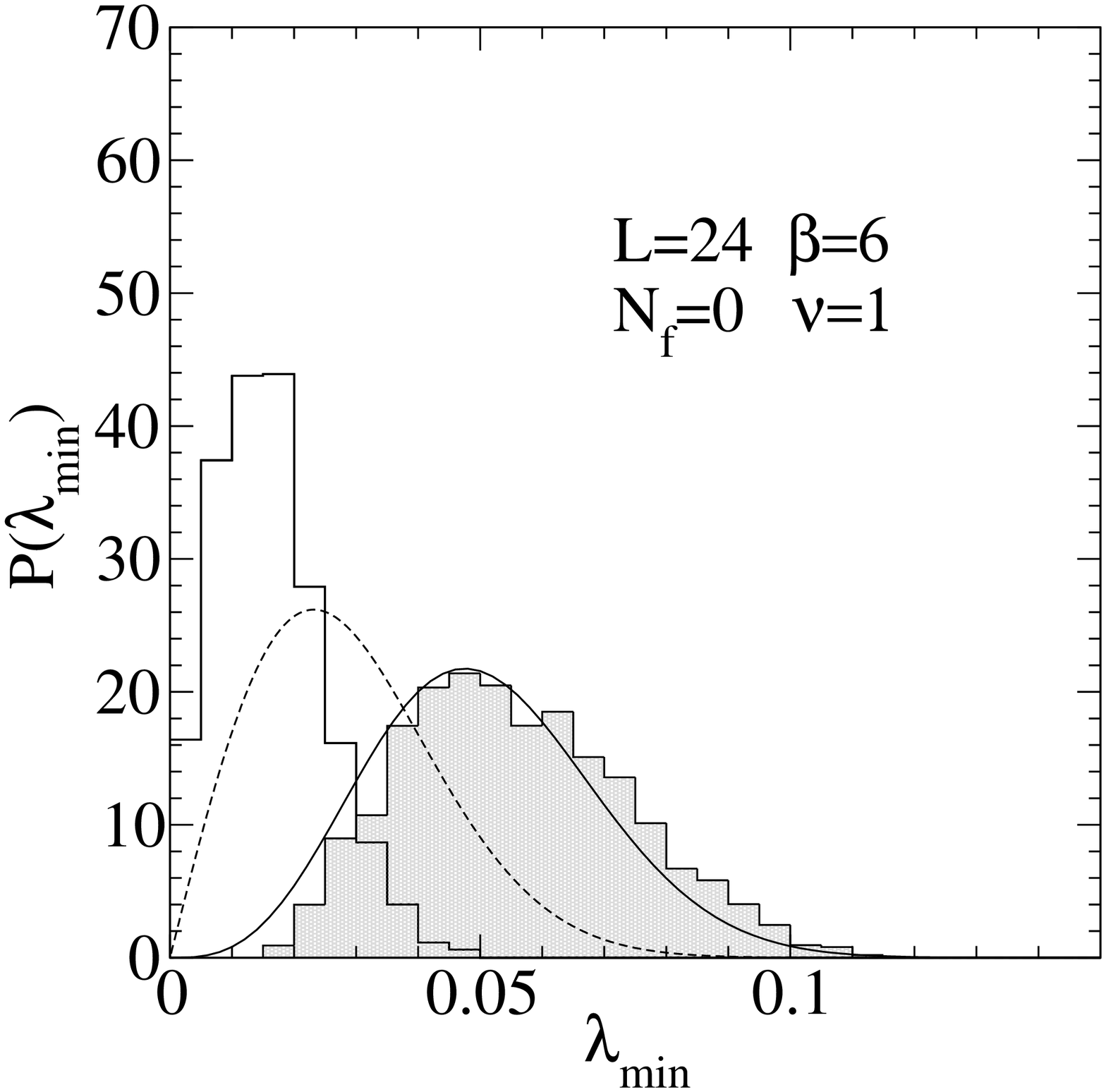,width=6.5 truecm}
\epsfig{file=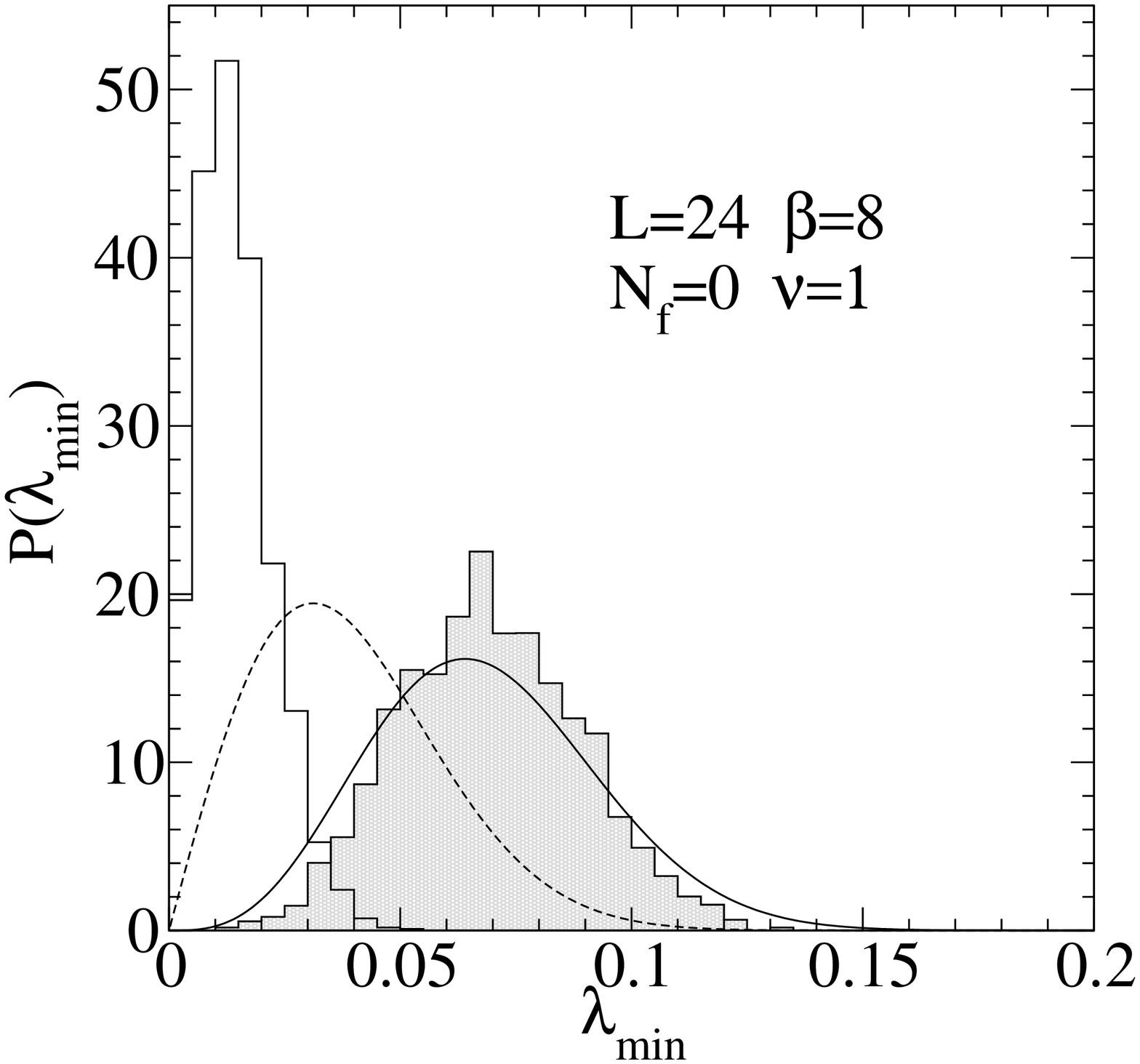,width=6.5 truecm}
\end{center}
\caption{\label{fig:min_top}
Distribution of the smallest and second
(shaded histogram) eigenvalue for $|\nu|=1$.
The continuous (dashed) curves represent the theoretical 
prediction of chUE for the smallest eigenvalue for $|\nu|=1$ ($\nu=0$)
based on the $\Sigma$-value from the fit to the $\nu=0$ data discussed
earlier.}
\end{figure}

It has been suggested \cite{Sm92} that the spectral density of the quenched
Schwinger model should  diverge exponentially with the volume for small
eigenvalues; our data do not allow us to draw any firm statement concerning 
this. We only observe that for $\beta=$2 and 4, $L=24$ and 32 give
consistent  values of $\Sigma$, while it is no more the case for  the larger
value of $\beta=6$, this indicating a slowing down of  the scaling behavior in
the weak coupling region.

\paragraph{Topological sector.}
Fig. \ref{fig:min_top} shows the distributions of the first and second
eigenvalue for configurations with $|\nu|=1$. Ideally, the  first eigenvalue
ought to be a zero mode, while the second eigenvalue  should follow the
predictions of chUE for the smallest eigenvalue  for $|\nu|=1$; these may be
obtained just using the values of $\Sigma$ found in the trivial sector in
the formulas for the smallest eigenvalue distribution with $|\nu|=1$  (the
continuous line in the Fig. \ref{fig:min_top}). 

What we see  instead is that for small $\beta$ the first  eigenvalue follows
the chUE statistics of the smallest eigenvalue in the trivial sector (dashed
curve in the figure).  Such a behavior was also observed for SU(3), d=4
\cite{DaHeNi99}. Increasing $\beta$, the first eigenvalue peaks closer and
closer to zero and the {\em second smallest} eigenvalue approaches the 
predictions of chRMT for the {\em smallest} eigenvalue for $|\nu|=1$
($\beta\ge 4$). We conclude that only towards the continuum limit  the zero
modes correlate clearly with the topological sector of the gauge
configuration.

In order to check the systematics of the possible misidentification, we
repeated the fits with chRMT shapes {\em without} selecting  the trivial
topological sector: even for the smallest value of $\beta$  at our disposal,
$\beta=2$, we always obtain considerably larger values of  $\chi^2/d.o.f.$
(data with asterisk in Table~\ref{tab:sigma})  and the fitting values of
$\Sigma$ are inconsistent with the determination in the trivial
sector.  We conclude, that even for moderate values of $\beta$,  the
identification of the topological sector is an essential ingredient for the
correct comparison with chRMT.

\paragraph{Re-weighting.}

\begin{figure}[t]
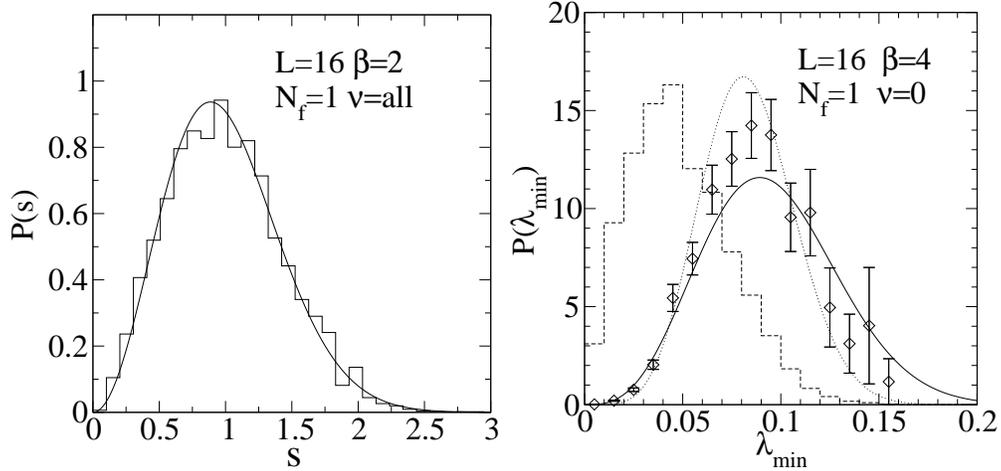

\begin{center}
\epsfig{file=figs/nndist_l16_b2_unq.eps,width=6.5 truecm}
\epsfig{file=figs/minev_l16_b4_unq_jk.eps,width=6.5 truecm}
\caption{\label{fig:min_unq}
Results for eigenvalues after re-weighting  with the fermion determinant,
corresponding to dynamical fermions.
Left: Level spacing distribution (unfolded variable), compared with the 
theoretical expectation for chUE\cite{GuMuWe98}.
Right: Distribution of the smallest eigenvalue (diamonds with error bars). 
The full and dotted curves are the best fits from chUE  with
$N_f=1$ and $N_f=2$ respectively, resulting in $\Sigma$ values  as given in
Table \ref{tab:sigma}. For comparison we report also the histogram of quenched
data (dashed line).}
\end{center}
\end{figure}

Of high interest in chRMT is the dynamics of fermions. In order to account for
dynamical fermions, we  re-weight  the quenched gauge configurations with the
fermion determinant; in the case of the statistics  $P(\lambda_{\rm min})$,
for example, this amounts to the  following redefinition:
\be
P(\lambda_{\rm min})=\frac{\sum_C \delta(\lambda_{\rm min}-\lambda_{\rm min}(C))\, {\rm det}
{\cal D}(C)}{\sum_C {\rm
det}{\cal D}(C)}\;\;,
\ee
where $\lambda_{\rm min}(C)$ denotes the smallest eigenvalue  for the
configuration $C$. The fluctuation of the determinant introduces 
additional fluctuations in the studied quantity (in our case the error
typically doubles),  which can be traced back to the  effective loss of
statistics of the gauge sample. Loss of statistics is also caused by the
depletion of the trivial sector (the one under study) on large lattices. 

As already discussed in \cite{FaHiLa99c}, in the quenched situation we find
perfect chUE behavior for the level spacing distribution  for all parameter
values studied. This observation persists also for the re-weighted, unquenched
data (e.g. Fig. \ref{fig:min_unq} left). For the level spacing we always find chUE
distribution. Such a behavior in a situation of vanishing fermion condensate
was found already in the Coulomb phase of compact four dimensional QED
\cite{BeMaPu99}.

In Fig. \ref{fig:min_unq} (right)  we report the distribution of the smallest
eigenvalue for $L=16$, $\beta=4$ and find it clearly different from the 
quenched distribution. As expected from the discussion of Sec.
\ref{sec:res_edge} neither the  $N_f=1$ chUE shape nor the $N_f=2$ succeed in
fitting the distribution. The two different fits result  in different values
of $\Sigma$  reported in Table \ref{tab:sigma} and Fig. \ref{fig:sigma},
right.  In both cases the values of $\chi^2/d.o.f.$ of the  fit are rather
large, confirming that chUE  does not give the right prediction. The fitted
values  of $\Sigma$ are well above zero.  As discussed, strictly speaking
$\Sigma$ ought to be zero, a different chRMT and universality class
applying. In the (true) one flavor situation the continuum condensate is 
$-\langle\bar{\psi}\psi\rangle_{\rm cont}/e \approx 0.15989$; this situation
is described by one copy of Overlap  or FP fermions \cite{FaHiLaWo99}.

\begin{table}[htb]
\small
\centering
\begin{tabular}{llllll}
\hline
 & $L$ & $\beta$ & quen. ($N_f=0$) & 
rew. ($N_f=1$)  	 & rew. ($N_f=2$)  \\
\hline 
 & 16 & 2 & 0.3282(31) [0.81] &0.2058(72) [0.56] & 0.363(14) [1.05] \\
$^{*)}$ &  16  & 2 &   0.3126(34) [3.26] & & \\
 &  24 & 2 & 0.3086(38) [0.65] & 0.120(22) [0.26] & 0.210(34) [0.28] \\
$^{*)}$  &  24 & 2 & 0.3264(36) [2.44] & & \\
 &  32 & 2 & 0.3289(67) [0.72] & 0.084(21) [1.41] & 0.149(26) [1.40] \\
$^{*)}$  &  32  & 2 & 0.3319(48) [3.84] &  &  \\
 &  16 & 4 & 0.1420(12) [0.83] &0.1256(22) [1.97] & 0.2026(54) [5.25]\\
 &  24 & 4 & 0.1741(37) [1.42] &0.1302(67) [0.79] & 0.184(12) [1.99] \\
 &  32 & 4 & 0.1860(40) [1.21] & 0.118(17) [0.43] & 0.178(14) [0.20] \\
 &  24 & 6 & 0.1047(18) [1.25] &0.0811(38) [4.83] &0.1391(38) [2.58] \\
 &  32 & 6 & 0.1271(25) [1.37] &0.0749(64) [3.93] &0.1273(64) [2.74] \\
 &  24 & 8 & 0.0787(18) [3.17] &0.0635(22) [3.85] &0.1110(27) [2.89] \\

\hline
\end{tabular}
\caption{\label{tab:sigma} 
Values of the fermion condensate obtained from comparison of lattice data
with chRMT (chUE). In the square brackets we indicate $\chi^2/d.o.f.$. Data
with asterisk are obtained  without selecting a topological sector.}
\end{table}

\section{Discussion}

We conclude that for a careful RMT analysis of the eigenvalue  spectrum of the
staggered Dirac operator the identification of the topological sector is
advisable. Towards the continuum limit the topologically non-trivial sectors 
exhibit an extra peak at small eigenvalues, which we attribute to the
``want-to-be'' zero modes. For a recent discussion on the
relevance of quasi-zero modes and their chirality in the context of staggered
fermions cf. \cite{VeKi97}.

We identified problems due to the general ambiguity of staggered  fermions
between the choice $N_f=1$ applying for finite lattice  spacing and
$N_f=2^{d/2}$ applying in the continuum limit.  For example, the quenched
microscopic spectral density clearly  deviates from chUE for
$\beta\ge 4$,  because the ``background'' estimate of $\Sigma$,
$\pi\rho(0)$, does not match with the ``microscopic'' one as can be obtained
e.g. from the smallest eigenvalue distribution.

Additional unexpected features come into play  because of our two-di\-men\-sio\-nal
context: the Mermin-Wagner-Coleman theorem implies a vanishing fermion
condensate with the result that chUE fails in the case of dynamical fermions.
This is indicated by our data, even if the statistics is not good enough to
draw firm conclusions. This scenario is quite different from that found in
\cite{FaHiLaWo99} where the Schwinger model with Ginsparg-Wilson fermions was
analyzed; there, genuine single-flavored fermions where described.

{\bf Acknowledgment:}

We want to thank C. Gattringer, I. Montvay and K. Splittorff for helpful 
discussions. F.F. acknowledges support from Schweizerischer Nationalfonds.

\end{document}